\documentclass[journal]{IEEEtran}

\usepackage{amsmath,amsfonts,amssymb,amscd,amsthm,dsfont}
\usepackage{graphicx,epsfig}
\usepackage{color}
\usepackage{mathtools}
\usepackage{bbm}
\usepackage{url}
\usepackage{balance}
\usepackage{epstopdf}
\usepackage{subfigure,enumitem}
\usepackage{cite,url,hyperref}
\usepackage{verbatim}
\usepackage{bm}
\usepackage{multicol}
\usepackage{ulem}
\usepackage{booktabs}
\usepackage{multirow}

\usepackage{caption}
\usepackage{lipsum}

\newcommand{\nn}{\nonumber}

\newtheorem{theorem}{\textbf{Theorem}}

\newtheorem{lemma}{\textbf{Lemma}}

\newtheorem{proposition}{\textbf{Proposition}}
\newtheorem{assumption}{Assumption}
\usepackage{color}

\allowdisplaybreaks
\begin{document}
\title{Data-Driven Quickest Change Detection in (Hidden) Markov Models}

\author{Qi Zhang\quad Zhongchang Sun \quad Luis C. Herrera \quad Shaofeng Zou
	\thanks{ This paper was presented in part at the 2023 IEEE International Conference on Acoustics, Speech, and Signal Processing \cite{zhang2023data} and the 2023 IEEE International Symposium on Information Theory \cite{zhang2023datadriven}.}
	\thanks{ The authors are with the Department of Electrical Engineering, University at Buffalo, Buffalo, NY 14228 USA (e-mail: 
  \href{mailto:qzhang48@buffalo.edu}
 {qzhang48@buffalo.edu},
 \href{mailto:zhongcha@buffalo.edu}{zhongcha@buffalo.edu},
 \href{mailto:lcherrer@buffalo.edu}
 {lcherrer@buffalo.edu},
 \href{mailto:szou3@buffalo.edu}{szou3@buffalo.edu}).}
}

%
\maketitle
\begin{abstract}
The paper investigates the problems of quickest change detection in Markov models and hidden Markov models (HMMs).  Sequential observations are taken from a (hidden) Markov model. At some unknown time, an event occurs in the system and changes the transition kernel of the Markov model and/or the emission probability of the HMM. The objective is to detect the change quickly, while controlling the average running length (ARL) to false alarm. The data-driven setting is studied, where no knowledge of the pre-, post-change distributions is available. Kernel-based data-driven algorithms are developed, which can be applied in the setting with continuous state, can be updated in a  recursive fashion, and are computationally efficient.   Lower bounds on the ARL  and upper bound on the worst-case average detection delay (WADD) are derived. The WADD is at most of the order of the logarithm of the ARL.  The algorithms are further numerically validated on two practical problems of fault detection in DC microgrid and photovoltaic systems.
\end{abstract}

\section{Introduction}
In the problem of quickest change detection (QCD), sequential observations are generated from a stochastic process. At some unknown time (change-point), an event occurs and causes the data-generating distribution to undergo a change. The objective is to quickly detect the change, while  controlling the false alarm rate. This QCD problem has been widely studied in the literature \cite{poor-hadj-qcd-book-2009,tartakovsky2014sequential,xie2021sequential,veeravalli2013quickest,basseville1993detection}. It models a wide range of applications, e.g., fault detection in DC microgrids \cite{gajula2021quickest},  quality control in online
manufacturing systems \cite{lai1995sequential}, spectrum monitoring in wireless communications \cite{lai2008cognitive},  early detection of epidemics \cite{baron2004early} and signal processing in  genetic area  \cite{shen2012change}. 

Most existing studies investigate the setting where pre- and post-change samples are independent and identically distributed (i.i.d.), respectively. 
 In \cite{pollak1985optimal,moustakides1986optimal,ritov1990decision}, the authors prove that the CuSum algorithm is optimal when the samples are i.i.d.. However, the i.i.d.\ assumption may not hold in practice, e.g., in photovoltaic systems \cite{chen2016quickest}, samples are not independent over time. In \cite{lai1998information, tartakovsky2005general,baron2006asymptotic,tartakovsky2017asymptotic}, the general QCD problem for the non-i.i.d.\ setting is studied,  where the approaches are model-based with known pre- and post-change distributions. In \cite{yakir1994optimal,polansky2007detecting,darkhovsky2011change}, the case where samples are generated from a Markov model is investigated. In these literature, both the pre- and post-change  distributions are  known exactly, and therefore those algorithms may not be applicable if such information is unavailable or the performance may degrade significantly if such information is inaccurate. In this paper, we first study QCD problem in Markov models under the data-driven setting, where neither
the pre- nor the post-change transition kernel is known.  We then focus on the more challenging hidden Markov models (HMMs), where  the pre- and 
post-change transition kernels and emission probability distributions
are unknown. 

\subsection{Related Works}
For Markov models, the data-driven setting where the pre- and the post-change transition kernels are unknown is studied in, e.g.,\cite{xian2016online,chen2022change}. 
QCD in Markov models with known pre-change and unknown post-change transition kernels is investigated in \cite{xian2016online}. The maximum likelihood estimate (MLE) of the unknown post-change transition kernel is employed to construct a generalized CuSum-type test. However, this method only works for finite state Markov chains or parameterized Markov transition kernels, which is not applicable to general problems with continuous state. QCD in Markov models with both pre- and post-change transition kernels unknown is studied in \cite{chen2022change}, where a kernel-based method is developed. A kernel method of maximum mean discrepancy (MMD) \cite{sriperumbudur2010hilbert} is employed to construct a CuSum-type test.  The kernel MMD method for data-driven QCD under the i.i.d.\ setting is also investigated in e.g., \cite{flynn2019change,li2019scan}. However, they focus on the case with i.i.d.\ samples, and thus their approach cannot be directly applied to our Markov models. 


In \cite{fuh2003sprt,fuh2004asymptotic,fuh2018asymptotic,fuh2015quickest,sun2023quickest},  QCD  in HMMs is studied.
All these studies investigate the model-based setting with both pre- and post-change information known.  However, the information may not be accurate or available in practice, especially when the change is due to some unexpected fault.
In \cite{fuh2003sprt}, the model-based setting with known pre- and post-change distributions for QCD in HMMs with a \textit{finite} state space is  investigated. The $L_1$-norm of products of Markov random matrices are employed to construct a CuSum test. This approach however only applies to  HMMs with a finite state space.
In \cite{fuh2004asymptotic}, the same problem is investigated. The ratio of the $L_1$-norms of products of Markov random matrices is employed to construct the log-likelihood in their Shiryayev–Roberts–Pollak test.
In \cite{fuh2015quickest}, a recursive algorithm is designed for the same problem. A quasi-generalized likelihood ratio is employed to update the algorithm in a recursive fashion. The proposed approach however only applies to HMMs with two states and does not generalize to problems with a continuous state space.  QCD in HMMs under Bayesian setting  is studied in \cite{fuh2018asymptotic}, which assumes the change point follows some prior distribution.
All the above studies investigate QCD in HMMs with a finite state space and need both the pre- and post-change transition kernels and emission probability distributions. 

In \cite{sun2023quickest}, QCD in auto-regressive (AR) model is studied, where observations are i.i.d. before the change and follow an AR model after the change. A computationally efficient algorithm is proposed. However, for the data-driven setting, only the lower bound on average running length (ARL) is provided. QCD in AR models is also studied in \cite{chen2016quickest}, where the likelihood ratio is approximated using a data-driven method. However, the authors did not provide any theoretical guarantee. 

\subsection{Major Contributions}

We develop kernel-based data-driven  algorithms for QCD in (hidden) Markov models. 
We keep a sample buffer of size $m$. Once the buffer is filled, we calculate the MMD between samples in the buffer and a reference batch of pre-change samples. We use this MMD to construct a CuSum-type test. Subsequently, we clear the buffer and continue sampling. The statistic has a negative drift, causing the CuSum to fluctuate around 0 before the change. After the change, it has a positive drift, leading the CuSum to exceed the threshold quickly.  We theoretically demonstrate that the lower bound on ARL increases exponentially with the threshold and that the upper bound on the worst-case average detection delay (WADD) is linear in the threshold for both Markov models and HMMs. Hence, the WADD is at most in the logarithm order of the ARL. This result matches with the result of QCD under the model-based setting. 
Combined with the universal lower bound on the WADD in \cite{lai1998information}, our algorithm is optimal at the order-level. The primary difficulty in our proof arises from the fact that the samples are generated from (hidden) Markov models, and thus it is essential to explicitly characterize the bias in the MMD estimate for the analysis of ARL and WADD. 

When this paper was in preparation, the first version of \cite{chen2022change} was posted where the ARL lower bound grows linearly with the threshold, which indicates overly frequent false alarms. Later \cite{chen2022change} provides an improved result that the ARL is at least exponentially in the threshold and the upper bound on WADD is linear in the threshold. However, the lower bound on ARL in \cite{chen2022change} requires the threshold to be sufficiently large. Here we do not need such assumption. In addition, \cite{chen2022change} focuses on Markov models. To the best of the authors' knowledge, we are the first to develop data-driven algorithms for QCD in HMMs with theoretical performance guarantees and order-level optimal performance. Our algorithm has a complexity of $\mathcal O(mn)$ for every $n$ samples, whereas the computational complexity in \cite{chen2022change} is $\mathcal O(m^2n)$.
Our simulation results also demonstrate that our algorithm has a smaller WADD for a given ARL, and is computationally efficient. 

We conduct extensive experiments on synthetic data and two practical fault detection problems in power systems.  The numerical results show that for fault detection in DC microgrids, our algorithm outperforms the algorithm proposed in \cite{chen2022change} in data-driven settings. For line-to-line fault detection in photovoltaic systems, our data-driven algorithm outperforms the model-based methods if applied with inaccurate system knowledge in \cite{gustafsson2000adaptive,liu2017sensor}.

This journal paper provides complete proofs for the results, which were omitted due to space limitation in the conference papers \cite{zhang2023data,zhang2023datadriven}. Moreover, we also add extensive experiments on practical applications of fault detection in DC microgrid and photovoltaic systems, and compare our algorithms with existing approaches.

\section{Problem Formulation}
\textit{Markov Models}: 
Consider a Markov chain $\{X_t\}_{t=1}^\infty$ defined on a probability space $(\mathcal X,\mathcal F, \mathbb P)$ with $
\mathbb P(X_t\in A|X_{t-1},\ldots,X_1)= \mathbb P(X_t\in A|X_{t-1})
$
for each $t\ge 1$ and $A\subseteq \mathcal X$. Define the transition kernel $ P: \mathcal X \to \mathcal P(\mathcal X)$ for the Markov chain $\{X_t\}_{t=1}^\infty$ 
, where $P(X_{t}|X_{t-1})$ is the  transition probability density of $X_{t}$ given $X_{t-1}$ and $\mathcal P(\mathcal X)$ denotes the probability simplex on $\mathcal X$.
The transition probability density changes at an unknown time $\tau$ from $P$ into $ Q: \mathcal X \to \mathcal P(\mathcal X)$. 

\textit{Hidden Markov Models}: The state $X_t$ of the Markov chain cannot be observed directly. Instead, we have access to an observable sequence  $\{ X_t'\}_{t=1}^\infty$ which is adjoined to the Markov chain $\{X_t\}_{t=1}^\infty$ such that $\{X_t,X'_t\}_{t=1}^\infty$ is a Markov chain. For any measurable set $A\subseteq \mathcal X$, the following condition holds:
\begin{align*}
\mathbb P(X_t\in A|X_{t-1},\ldots,X_1,X'_{t-1},\ldots,X'_1)&= P(X_t\in A|X_{t-1}),\\
\mathbb P(X'_t\in A|X_{t},\ldots,X_1,X'_{t-1},\ldots,X'_1)&=P'(X'_t\in A|X_{t}).
\end{align*}
Let $ P': \mathcal X \to \mathcal P(\mathcal X)$ be the emission probability. 
At some unknown time $\tau$, the transition kernel changes from $P$ to $Q$ and/or the emission probability changes from $P'$ to $ Q'$. 

The objective is to detect the change quickly subject to false alarm constraints.  In this paper, we investigate the data-driven setting, where  $P, P', Q, Q' $ are \textit{unknown}.  

 Let $\hat T$ be a stopping time and let $\mathbb P_{\tau}$ ($\mathbb E_{\tau}$) be the probability measure (expectation) when the change happens at time $\tau$. Let $\mathbb P_{\infty}$  ($\mathbb E_{\infty}$) be the probability measure (expectation) if no change happens.  There is a reference sequence of samples $\{Y_t\}_{t=1}^\infty$ from transition kernel $ P$ \cite{chen2022change,li2019scan,flynn2019change}. Use $\mathcal{F}_t$ to denote the $\sigma$-field generated by $\{X_i,Y_i\}_{i=1}^t$. For HMMs, there is a reference sequence of observations $\{Y_t'\}_{t=1}^\infty$ generated from a HMM with transition kernel $P$ and emission probability $P'$. Denote the hidden state by $Y_t$. Also use $\mathcal{F}_t$ to denote the $\sigma$-field generated by $\{X_i,X_i',Y_i,Y_i'\}_{i=1}^t$.
Define the ARL and the WADD for $\hat{T}$:
\begin{flalign}
      &\text{ARL}(\hat T)=\mathbb E_{\infty}[\hat T],\\
      &\text{WADD}(\hat T)=\sup_{\tau \ge 1}  \mathbb  E_{\tau}[(\hat T-\tau)^+|\mathcal F_{\tau-1}].
\end{flalign}
Here the ARL quantifies the frequency of false alarms, and the WADD demonstrates the number of samples required to trigger an alarm.
The objective is to minimize the WADD under the ARL constraint:
\begin{flalign}
\min_{\hat{T}} \text{WADD}(\hat T),\text{s.t.} \text{ ARL}(\hat T)\ge \psi,\label{eq:goal}
\end{flalign}
where $\psi>0$ is some pre-specified constant.
We assume that the Markov chains with transition kernels $P$ and $Q$ are uniformly ergodic \cite{chen2022change,xian2016online}, i.e., for any measurable  $A\subseteq\mathcal X$ and $x\in \mathcal X$
    \begin{align}
        |\mathbb P_\infty(X_{t+i} \in A|X_i=x)-\pi_P(A)|\le R_P\lambda^t_P,\\
         |\mathbb P_1(X_{t+i} \in A|X_i=x)-\pi_Q(A)|\le R_Q\lambda^t_Q,
    \end{align}
    where $\pi_P$ and $\pi_Q$ denote the invariant distributions for Markov models with transition kernels $P$ and $Q$, and $0<R_P,R_Q<\infty$, $0<\lambda_P, \lambda_Q<1$ are some constants.
It can be demonstrated that under $\mathbb P_\infty$, the invariant distribution of $\{X_t,X'_t\}_{t=1}^\infty$ is $\pi_P(x) P'(x'|x)$, and under $\mathbb P_1$, the  invariant distribution is $\pi_Q(x)  Q'(x'|x)$. Let
$
    {\pi}_P'(x')=\int_{\mathcal X}\pi_P(dx) P'(x'|x),$ and $
    {\pi}_Q'(x')=\int_{\mathcal X}\pi_Q(dx) Q'(x'|x)
$
be the marginal distribution of the observation under the invariant distribution by ${\pi}_P'$ and ${\pi}_Q'$, respectively.
\subsection{Maximum Mean Discrepancy}\label{sec:mmd}
In this section, we provide a brief introduction to kernel mean embedding and MMD \cite{berlinet2011reproducing,sriperumbudur2010hilbert}. Consider a  positive definite kernel function $k$: $\mathcal{X} \times \mathcal{X} \to \mathbb{R}$, in a reproducing kernel Hilbert space (RKHS) denoted by $\mathcal H_k$. Let $\langle \cdot,\cdot\rangle_{\mathcal{H}_k}$ be the inner product in the RKHS. For any $x,y\in\mathcal X$, we have $ k(x,\cdot) \in \mathcal H_k$ and $k(x,y)=\langle k(x,\cdot),k(y,\cdot)\rangle_{\mathcal{H}_k}.$  Following the reproducing property,  any function $f \in \mathcal{H}_k$, $f(x) = \langle f,k(x,\cdot)\rangle_{\mathcal{H}_k}$.  In this paper, we study a bounded kernel function $k$, i.e., $0\le k(x,y)\le1$. Let $\mu_F=\mathbb E_{X\sim F}[k(X,\cdot)]$ be the kernel mean embedding of a probability distribution $F$ . For a characteristic kernel  $k$, $\mu_{F}=\mu_{G}$ if and only if the probability $G$ and $F$ are identical\cite{sriperumbudur2010hilbert}. 
The definition of MMD between distribution $F$ and $G$ is as follows:
\begin{equation}
    D(F,G)=\sup_{f :\|f\|_{\mathcal{H}_k}\le 1 }\Big |\mathbb E_{X\sim F}[f(X)]-\mathbb E_{Y\sim G}[f(Y)]\Big|.
\end{equation}
The squared MMD can be written equivalently as
\cite{gretton2012kernel}:
\begin{flalign}\label{eq:mmd}
    D^2(F,G)&=\mathbb E_{X,\bar X\sim F}[k(X,\bar X)]+\mathbb E_{Y,\bar Y\sim G}[k(Y,\bar Y)]\nonumber\\
    &-2\mathbb E_{X\sim F,Y\sim G}[k(X,Y)].
\end{flalign}
\section{Markov Models}\label{sec:markov}
In this section, we present our results for Markov models.
Markov chains with distinct transition kernels may generate identical stationary distributions \cite{chen2022change}.  As a result, a straightforward extension of existing QCD methods designed for the i.i.d.\ setting \cite{flynn2019change,li2019scan}, which estimates the  MMD between stationary distributions $\pi_P$ and $\pi_Q$, may not be work even  $P\neq Q$.  To address this issue, we utilize the second-order Markov chain \cite{chen2022change}. We define the product $\sigma$-algebra on $\mathcal{X} \times \mathcal{X}$, which is generated by the collection of all
measurable rectangles as follows:
$
       \mathcal{F} \otimes \mathcal{F} =\sigma\{A \times B: A \subseteq \mathcal X, B\subseteq \mathcal X\}. 
$
Define the second-order probability measure $\widetilde{\pi}_P$ on measurable space $(\mathcal{X} \times \mathcal{X},\mathcal{F} \otimes \mathcal{F})$  as follows:
$
    \widetilde{\pi}_P(A \otimes B) =
    \int_{A} \pi_P (dx) P(X_{i+1}\in B|X_{i}=x),
$
where $A$ and $B$ are elements of the $\mathcal{F}$ $\sigma$-algebra.
According to Lemma 1 in \cite{chen2022change}, if the transition kernels are distinct, the second-order probability measures are distinct. 
The kernel function and MMD discussed in Section \ref{sec:mmd} can be straightforwardly extended to the space of $\widetilde{\mathcal{X}} =\mathcal{X} \times \mathcal{X}$. 

It can be easily shown that 
under $\mathbb{P}_{\infty}$ ($\mathbb{P}_{1}$), the second-order Markov chain $\{\widetilde{X}_t=(X_t,X_{t+1})\}_{t=1}^\infty$ is uniformly ergodic.
We then construct the test statistic and the stopping rule using the second-order Markov chain. Our approach is to estimate the MMD between the invariant measures of the second-order Markov chains, and employ this estimate to construct a CuSum-type test. 

As a first step, we divide the samples into non-overlapping blocks of $m$. Let $\widetilde{X}_t=(X_t,X_{t+1})$, $\widetilde{Y}_t=(Y_t,Y_{t+1})$ be the second-order Markov chains. For the observation sequence$\{\widetilde{X}_t\}_{t=1}^\infty$, the pre-change stationary distribution is $ \widetilde{\pi}_P$ and the post-change stationary distribution is $\widetilde{\pi}_Q$. For the reference sequence $\{\widetilde{Y}_t\}_{t=1}^\infty$, the stationary distribution is $ \widetilde{\pi}_P$. For the $t$-th block, $t=0,1,2,\ldots$,
define the empirical measure of  $\{\widetilde{X}_{mt+i}\}_{i=1}^{m-1}$ as $F^t_{\widetilde X}=\frac{1}{m-1}\sum_{i=1}^{m-1} \delta_{\widetilde{X}_{mt+i}},$
where $\delta_{\widetilde{X}_{i}}$ is the Dirac measure. Similarly, we can define $ F^t_{\widetilde Y}$.
The MMD $D(F^t_{\widetilde X},F^t_{\widetilde Y})$ can then be computed using \eqref{eq:mmd}.
Define $S_t=D(F^t_{\widetilde X},F^t_{\widetilde Y})-\sigma,$
 where $0<\sigma<D(\widetilde{\pi}_P, \widetilde{\pi}_Q)$ is a positive constant that will be determined later.
 Then, our stopping time is defined as
 \begin{equation}
     T(c)=\inf  \Bigg\{ mt+t: \max_{0\le i\le t } \sum_{j=i}^tS_j>c \Bigg\} \label{eq:define},
 \end{equation}
  where $c>0$ is  a predetermined threshold.
  
Our algorithm can be updated in a recursive fashion:
$
  \max_{0\le i\le t }\sum_{j=i}^tS_j=\max\Big\{0,\max_{0\le i\le t-1 } \sum_{j=i}^{t-1}S_j+S_t\Big\}.
$ Furthermore, for every block with size $m$ the computational complexity for MMD is $\mathcal O(m^2)$. At time $n$, there are $\lfloor\frac{n}{m}\rfloor$ non-overlapping blocks in total. Thus, the total computational complexity up to time $n$ is $\mathcal O(mn)$. In contrast, the algorithm in \cite{chen2022change}  uses overlapping blocks and has an overall computational complexity of  $\mathcal O(m^2n)$. 

Intuitively, when $m$ is large, we have that $D(F^t_{\widetilde X},F^t_{\widetilde Y})\approx D( \widetilde{\pi}_P, \widetilde{\pi}_P)= 0$ before the change and $D(F^t_{\widetilde X},F^t_{\widetilde Y})\approx D(\widetilde{\pi}_P, \widetilde{\pi}_Q)>0$ after the change. For $0<\sigma<D(\widetilde{\pi}_P, \widetilde{\pi}_Q)$, the test statistic has a negative drift before the change, causing the CuSum to fluctuate around $0$, while after the change, it has a positive drift, leading the CuSum to exceed the threshold quickly.

We present the upper bound on the WADD of  \eqref{eq:define} in the following theorem.
Define 
$
    a_P=\sqrt{\frac{2-2\lambda_P+4R_P}{(m-1)(1-\lambda_P)}},  a_Q=\sqrt{\frac{2-2\lambda_Q+4R_Q}{(m-1)(1-\lambda_Q)}},
    a=a_P+a_Q,$ and $ 
    d=D(\widetilde{\pi}_P, \widetilde{\pi}_Q)-\sigma.
$
\begin{theorem}\label{thm:add}
The WADD for the stopping time in \eqref{eq:define} can be upper bounded as follows:
\begin{flalign}
 \textup{W}  \textup{ADD}(T(c)) 
  \le&  \frac{2\sqrt{ad}mc}{(a-d)^2}
  + \frac{(a+d)mc}{(a-d)^2}+2m+\frac{a+\sqrt{ad}}{d-a}m\nn\\
  =&\mathcal O(mc).\label{eq:wadd}
\end{flalign}
\end{theorem}
\begin{proof}
Let $\tau'$ be the index of the last sample within the block where the change occurs. Thus, samples after $\tau'$ are generated from  transition kernel $Q$. Moreover, we have $(T(c)-\tau)^+\le (T(c)-\tau')^++m$. 
Let $$\xi=\\\\ \frac{a_P+a_Q+\sqrt{(D(\widetilde{\pi}_P, \widetilde{\pi}_Q)-\sigma)(a_P+a_Q)}}{D(\widetilde{\pi}_P, \widetilde{\pi}_Q)-\sigma-a_P-a_Q}.$$ It can be shown that $ D(\widetilde{\pi}_P, \widetilde{\pi}_Q)-\sigma-\frac{\xi+1}{\xi}(a_P+a_Q) >0.$
Denote by  $n_c=\Big\lceil\frac{(\xi+1)c}{D(\widetilde{\pi}_P, \widetilde{\pi}_Q)-\sigma}\Big\rceil$. Firstly, for any $\tau$ and $\mathcal F_{\tau-1}$ we have that
\begin{flalign}
     & \mathbb E_{\tau}\Bigg[\frac{(T(c)-\tau')^+}{mn_c}\Big|\mathcal F_{\tau-1}\Bigg]\le  \sum_{j=0}^{\infty}\mathbb P_{\tau}\Bigg(\frac{T(c)-\tau'}{mn_c}\ge j\Big|\mathcal F_{\tau-1}\Bigg).\nonumber
 \end{flalign}
According to the definition of $T(c)$, it can be shown that
\begin{flalign}
   \mathbb P&_{\tau}\Bigg(\frac{T(c)-\tau'}{mn_c}\ge j\Big|\mathcal F_{\tau-1}\Bigg)\nonumber\\
\le& \mathbb P_{\tau}\Bigg(\sum_{i=(j-1)n_c+1}^{jn_c}S_i<c\Big|\mathcal F_{\tau-1},\frac{T(c)-\tau'}{mn_c}\ge j-1\Bigg)\nonumber\\
&\times \mathbb P_{\tau}\Bigg(\frac{T(c)-\tau'}{mn_c}\ge j-1\Big|\mathcal F_{\tau-1}\Bigg).  \label{eq:prob2}
\end{flalign}
 It then suffices to bound  $ \mathbb P_{\tau}\Big(\sum_{i=(j-1)n_c+1}^{jn_c}S_i<c\Big|\mathcal F_{\tau-1},$
 $\frac{T(c)-\tau'}{mn_c}\ge j-1\Big)$ and to employ \eqref{eq:prob2} iteratively.
By the triangle inequality and Markov inequality, we have that
\begin{flalign}
\mathbb P&_{\tau}\Bigg (\sum_{i=(j-1)n_c+1}^{jn_c}S_i<c\Bigg |\mathcal F_{\tau-1},\frac{T(c)-\tau'}{mn_c}\ge j-1\Bigg)\nonumber\\
\le& \mathbb P_{\tau}\Bigg (\sum_{i=(j-1)n_c+1}^{jn_c}(D(\widetilde{\pi}_P, \widetilde{\pi}_Q)-\sigma -D( F_{\widetilde X}^i, \widetilde{\pi}_Q)\nonumber\\
&-D( F_{\widetilde Y}^i, \widetilde{\pi}_P))<c\Bigg |\mathcal F_{\tau-1},\frac{T(c)-\tau'}{mn_c}\ge j-1\Bigg )\nonumber\\
      \le & \mathbb P_{\tau}\Bigg (\sum_{i=(j-1)n_c+1}^{jn_c}(D( F_{\widetilde X}^i, \widetilde{\pi}_Q)+D( F_{\widetilde Y}^i, \widetilde{\pi}_P)) \nonumber\\
      &\ge n_c(D( \widetilde{\pi}_P, \widetilde{\pi}_Q)-\sigma)-c\Bigg |\mathcal F_{\tau-1},\frac{T(c)-\tau'}{mn_c}\ge j-1\Bigg )\nonumber\\
      \le & \sum_{i=(j-1)n_c+1}^{jn_c}\frac{ \mathbb  E_{\tau}\Big[D( F_{\widetilde X}^i, \widetilde{\pi}_Q)\Big|\mathcal F_{\tau-1},\frac{T(c)-\tau'}{mn_c}\ge j-1\Big]}{n_c(D( \widetilde{\pi}_P, \widetilde{\pi}_Q)-\sigma)-c}  \nonumber\\
      &+\frac{ \mathbb  E_{\tau}\Big[D( F_{\widetilde Y}^i, \widetilde{\pi}_P)\Big|\mathcal F_{\tau-1},\frac{T(c)-\tau'}{mn_c}\ge j-1\Big]}{n_c(D( \widetilde{\pi}_P, \widetilde{\pi}_Q)-\sigma)-c}\nonumber\\
      \le & \frac{a_P+a_Q}{D( \widetilde{\pi}_P, \widetilde{\pi}_Q)-\sigma}\left(1+\frac{1}{\xi}\right).\label{eq:prop1}
\end{flalign}
The last inequality in \eqref{eq:prop1} is from Proposition \ref{proposition:mean}, which can be found in Appendix \ref{proof:proposition 1}. Let $\delta=\frac{a_P+a_Q}{D( \widetilde{\pi}_P, \widetilde{\pi}_Q)-\sigma}\left(1+\frac{1}{\xi}\right)$, for any $\tau$ and $\mathcal F_{\tau-1}$ we get $ \mathbb P_{\tau}\Big(\frac{T(c)-\tau'}{mn_c}\ge j|\mathcal F_{\tau-1}\Big)\le \delta^j$ and thus $ \mathbb E_{\tau}\Big[\frac{(T(c)-\tau')^+}{mn_c}\Big|\mathcal F_{\tau-1}\Big]< (1-\delta)^{-1}$. Therefore,
\begin{flalign}
      &\mathbb E_{\tau}\Big[(T(c)-\tau')^+\Big|\mathcal F_{\tau-1}\Big] \le (1-\delta)^{-1}mn_c\nonumber\\
      &\le (1-\delta)^{-1}m\Bigg(\frac{(\xi+1)c}{D(\widetilde{\pi}_P, \widetilde{\pi}_Q)-\sigma}+1\Bigg).
\end{flalign}
This concludes the proof.
\end{proof}
Note that ${a}_P, a_{Q} = \mathcal O\Big(\frac{1}{\sqrt{m}}\Big)$. Thus, the upper bound increases linearly with the threshold $c$ and the batch size $m$.

We now derive a lower bound on the ARL, which increases exponentially with the threshold $c$.  
Let $h=\sigma-2a_P$. Due to the fact $a_P=\mathcal O\left(\frac{1}{\sqrt{m}}\right)$, there always exists  a large  $m$ such that $h>0$. Define a function $\phi(q)= \frac{\sqrt{\pi}}{\sqrt{\Gamma}}q \exp \Big({-qh+\frac{q^2}{4\Gamma}}\Big)$. 
We choose a constant $q>0$ such that $\phi(q) \le1$. 
Note that $\phi(q)$ is a continuous function of $q$, with $\phi(0)=0$ and $\phi(q) \to \infty$ as $q \to \infty$. Thus, there always exists such a value for $q$.

\begin{theorem}\label{thm:arl}
The ARL of $T(c)$ in \eqref{eq:define} can be lower bounded exponentially in the threshold $c$:
$
    \textup{ARL}(T(c))\ge m\exp(qc).
$
\end{theorem}

 \begin{proof}
Let $
   \Delta_1=\inf \Big\{t:\sum_{i=0}^t S_i<0 \Big\},\nonumber
 $ and 
 $
     \Delta_{r+1}=\inf \Big\{t>\Delta_r:\sum_{i=\Delta_r+1}^t\nonumber S_i<0 \Big\}.
 $ 
 We first show that for any $i$,
$
    \mathbb E_{\infty}\Big[\exp \left({qS_i}\right)|\mathcal{F}_{im}\Big]\le1.
$
Since $q>0$, we have that for any $l>0$,
\begin{flalign}
    \mathbb P&_{\infty}\big (\exp({qS_i})>l\big |\mathcal{F}_{im}\big )\nonumber\\
    =&\mathbb P_{\infty}\Bigg (D( F_{\widetilde X}^i, F_{\widetilde Y}^i)>\sigma+\frac{\log l}{q}\big |\mathcal{F}_{im}\Bigg )\nonumber\\
    \overset{(a)}{\le} & \mathbb P_{\infty}\Bigg (D(F_{\widetilde X}^i, \widetilde{\pi}_P)+D( F_{\widetilde Y}^i, \widetilde{\pi}_P)>2a_P \nonumber\\
    &+\frac{1}{\sqrt{\Gamma}}\left(\sigma+\frac{\log l}{q}-2a_P\right)\sqrt{\Gamma}\Big|\mathcal{F}_{im}\Bigg)\nonumber\\
    \overset{(b)}{\le} & \exp\Bigg({-\Gamma\left(\sigma+\frac{1}{q} \log l-2a_P\right)^2}\Bigg),\nonumber
\end{flalign}
where (a) is from the triangle inequality, and (b) is from Proposition \ref{prop:prob}, which
will be given later. Therefore, we have that
\begin{flalign}
     \mathbb E&_{\infty}\Big[\exp({qS_i})|\mathcal{F}_{im}\Big]
     = \int_0^\infty \mathbb P_{\infty}(\exp({qS_i})>l|\mathcal{F}_{im})dl\nonumber\\
     \le & \int_0^\infty \exp\Bigg({-\Gamma\left(\sigma+\frac{1}{q} \log l-2a_P\right)^2}\Bigg)dl.\nonumber
\end{flalign}
Let $\log l=\nu$, we get $dl=\exp(\nu)d\nu$. Therefore,
\begin{flalign}
      \mathbb E&_{\infty}\Big[\exp({qS_i})|\mathcal{F}_{im}\Big]\nonumber\\
      \le &\int_{-\infty}^\infty  \exp\Bigg({-\Gamma\Bigg(h+\frac{\nu}{q} \Bigg)^2}\Bigg)\exp(\nu)d\nu\nonumber\\
      =& \frac{\sqrt{\pi}}{\sqrt{\Gamma}} q \exp\Bigg({-qh+\frac{q^2}{4\Gamma}}\Bigg) 
      \le 1.\label{ineq:mean1}
\end{flalign}
 Define $R$ as  
 $
 R=\inf \Big\{r \ge 0: \Delta_r < \infty$  and $
 \sum_{i=\Delta_r+1}^t S_i \ge c$ for some $t>\Delta_r\Big\}.
 $
 Following the definition of $T(c)$, it can be shown that
 \begin{flalign}
 \mathbb E_{\infty}[T(c)]\ge m \mathbb E_{\infty}[R]= m\sum_{r=0}^\infty \mathbb  P_{\infty}(R>r).
 \end{flalign}
 We then bound $\mathbb  P_{\infty}(R>r)$.
For any $j\ge1$, we have
\begin{flalign}
\mathbb E&_{\infty}\Bigg[\exp\left(q\sum_{i=\Delta_r+1}^{\Delta_r+j} S_i\right)\Bigg|\mathcal{F}_{({\Delta_r+j})m}\Bigg]\nonumber\\
=&\exp\left(q\sum_{i=\Delta_r+1}^{{\Delta_r+j-1}} S_i\right)\mathbb E_{\infty}\Bigg[\exp\Big(qS_{({\Delta_r+j})}\Big)\Bigg|\mathcal{F}_{({\Delta_r+j})m}\Bigg] \nonumber\\
\le& \exp\left(q\sum_{i=\Delta_r+1}^{{\Delta_r+j}-1}S_i\right), \nonumber
\end{flalign}
where the last inequality is from \eqref{ineq:mean1}. Therefore $\exp\Big(q\\\ \sum_{i=\Delta_r+1}^{{\Delta_r+j}} S_i\Big)$ is a non-negative
supermartingale. Hence
\begin{flalign}
    \mathbb P&_{\infty}\Bigg(\max_{n\ge \Delta_r+1}\sum_{i=\Delta_r+1}^n S_i\ge c|\mathcal F_{\Delta_rm}\Bigg) \nonumber\\
    =&\mathbb P_{\infty}\Bigg(\max_{n\ge \Delta_r+1}\exp\left({q\sum_{i=\Delta_r+1}^n S_i}\right)\ge \exp({qc})|\mathcal F_{\Delta_rm}\Bigg)\nonumber \\
    \le &\mathbb E_{\infty}\Bigg[\exp(q S_{\Delta_r+1})|\mathcal F_{\Delta_rm}\Bigg]/\exp({qc})\nonumber\\
    \le & \exp({-qc}), \nonumber
\end{flalign}
where the first inequality is from  \cite[Lemma A.2]{flynn2019change} and  \cite[Theorem 7.9.2]{gallager2011discrete}.
Then we can get
\begin{flalign}
\mathbb P&_{\infty}(R\ge r+1|\mathcal F_{\Delta_rm})\nonumber\\
=& 1-\mathbb P_{\infty}\Bigg(\max_{n\ge \Delta_r+1} \sum_{i=\Delta_r+1}^n S_i\ge c|\mathcal F_{\Delta_rm}\Bigg)\nonumber\\
\ge& 1-\exp(-qc). \nonumber
\end{flalign}
We then have that
\begin{flalign}
   \mathbb P_{\infty}(R>r)=& \mathbb E_{\infty}\left[\mathbb P_{\infty}(R\ge r+1|\mathcal F_{\Delta_rm})\mathbbm 1_{\{R\ge r\}}\right]\nonumber\\
    \ge &(1-\exp({-qc}))\mathbb  P_{\infty}(R>r-1)\nonumber,
\end{flalign}
where the indicator function $\mathbbm 1_{\{R\ge r\}}=1$ if $R\ge r$, otherwise  $\mathbbm 1_{\{R\ge r\}}=0$.
 This further suggests that
 \begin{flalign}
     \mathbb E_{\infty}[R]= \sum_{r=0}^\infty \mathbb  P_{\infty}(R>r) 
     \ge \sum_{r=0}^\infty (1-\exp({-qc}))^r=\exp({qc}). \nonumber
 \end{flalign}
 This concludes the proof.
 \end{proof}
It is worth noting that $q$ and $m$ are independent of $c$. Therefore, the lower bound on ARL increases exponentially with the threshold $c$. 
For any stopping time with $\text{ARL}\ge \psi$, the detection delay is at least $\mathcal O(\log(\psi))$ (according to the universal lower bound result in \cite{lai1998information}). By Theorem \ref{thm:arl},  we choose the threshold $c=\frac{\log(\psi)-\log(m)}{q}$ to guarantee the false alarm constraint. By Theorem \ref{thm:add}, this further implies that our algorithm attains a detection delay of $\mathcal O(\log(\psi))$. This result matches with (order-level) the universal lower bound in \cite{lai1998information} for the general non-i.i.d.\ setting. 

\section{Hidden Markov Models}
In this section, we present our results for HMMs.
Similar to Markov models,  for the HMMs, with distinct transition kernels and and/or emission probability distributions, their stationary distributions and/or the induced marginal distributions of the observations may remain identical.

We generalize the idea of using second-order Markov chain in Section \ref{sec:markov} to the HMMs and explore higher-order HMMs. Denote the $i$-th order HMM by:
$$\Big\{\big((X_t,X_t'),(X_{t+1},X_{t+1}'),\ldots,(X_{t+i-1},X_{t+i-1}')\big)\Big\}_{t=1}^\infty,$$
and the corresponding $i$-th order observation sequence:
$\{(X_t',X_{t+1}',\ldots,X_{t+i-1}')\}_{t=1}^\infty.$
It is worth noting that the $i$-th order HMM is an HMM, and remains uniformly ergodic. 
To simplify notation, let $\pi_P'$ and $\pi_Q'$ be
the marginal distributions of the $i$-th order observation under the corresponding stationary distributions.

For simplicity of the presentation, we focus on the case with $i=2$, and introduce the following assumption. This assumption guarantees that $\pi_P'(X'_t,X_{t+1}')$ and $\pi_Q'(X'_t,X_{t+1}')$ are distinct (as shown in Lemma \ref{assume} in Appendix \ref{proof:l1}), thus ensuring the detectability of the change.
Utilizing a higher order model will lead to a relaxed assumption than Assumption \ref{a3}, although it will introduce additional computational cost. Generalization to a higher order model is straightforward.
\begin{assumption}\label{a3}
There exist $A,B\subseteq \mathcal X$ such that $$\pi_P'(X'_{t+1}\in A|X_{t}'\in B)\neq \pi_Q'(X'_{t+1}\in A|X_{t}'\in B).$$   
\end{assumption}

Let $\widetilde{X}_t'=\big(X'_t,X'_{t+1}\big)$ be the second-order observation and $\widetilde{X}_t=\big(X_t,X_{t+1}\big)$ be the second-order hidden state. 
%
Based on Lemma \ref{assume} and Assumption \ref{a3}, it can be shown that $ \widetilde{\pi}_P'(\widetilde{X}_t')\ne  \widetilde{\pi}_Q'(\widetilde{X}_t')$. 
Let  $\widetilde{Y}_t'=\big(Y_t',Y'_{t+1}\big)$ denote the observation at time $t$ of the second-order reference sequence.  We partition the samples into non-overlapping blocks of size $m$. For the $t$-th observation block, where $t=0,1,2,\ldots$, we obtain $m-1$ second-order samples and the empirical distribution for this block is 
$F^t_{\widetilde X'}=\frac{1}{m-1}\sum_{i=1}^{m-1} \delta_{\widetilde{X}'_{mt+i}}$. Similarly, for the  $t$-th reference block, we have $F^t_{\widetilde Y'}=\frac{1}{m-1}\sum_{i=1}^{m-1} \delta_{\widetilde{Y}'_{mt+i}}.$ We then can calculate the MMD between these two empirical distributions using \eqref{eq:mmd}.

We define our test statistic as $S_t'=D\Big(F^t_{\widetilde {X}'},F^t_{\widetilde {Y}'}\Big)-\sigma'.$ 
 The offset $\sigma'$ is to ensure that the test statistic $S_t'$ has a negative drift prior to the change and a positive drift after the change.
 Then, we define our stopping time as
 \begin{equation}
     T(c)=\inf  \Bigg\{ mt+t: \max_{0\le i\le t } \sum_{j=i}^tS_j'>c \Bigg\} \label{eq:define2}.
 \end{equation}
This algorithm can be updated recursively and the computational complexity for $n$ samples is $\mathcal O(mn)$ (see Section \ref{sec:markov}).
 
Intuitively, when $m$ is large, we have that $D\Big(F^t_{\widetilde {X}'},F^t_{\widetilde {Y}'}\Big)\approx  0$ prior to the change and $D\Big(F^t_{\widetilde {X}'},F^t_{\widetilde {Y}'}\Big)\approx D\left( \widetilde{\pi}_P', \widetilde{\pi}_Q'\right)>0$ after the change. Thus, by selecting a offset $0<\sigma'<D\left( \widetilde{\pi}_P', \widetilde{\pi}_Q'\right)$, the test statistic has a negative drift before the change while after the change, it has a positive drift.
Let $d'=D( \pi_P', \pi_Q')-\sigma'$. 
We then provide the upper bound on the WADD.
\begin{theorem}\label{add2}
The WADD for the stopping time in \eqref{eq:define2} can be upper bounded as follows:
\begin{flalign}
 \textup{W} & \textup{ADD}(T(c)) \nn\\
  \le&  \frac{2\sqrt{ad'}mc}{(a-d')^2}
  + \frac{(a+d')mc}{(a-d')^2}+2m+\frac{a+\sqrt{ad'}}{d'-a}m\nn\\
  =&\mathcal O(mc).\label{eq:wadd2}
\end{flalign}
\end{theorem}
The proof is similar to Theorem \ref{thm:add}. An important step is to show that the expectation of the MMD between the marginal distribution of the stationary distribution $\pi'_P$ and the empirical probability $F^t_{\widetilde X'}$ under $\mathbb P_{\infty}$ and the MMD between $\pi'_Q$ and $F^t_{\widetilde X'}$ under $\mathbb P_{1}$ is upper bounded, which is shown in Proposition \ref{proposition:mean2} (see Appendix \ref{proof:lemma3}). 



We now provide a lower bound on the ARL of $T(c)$ in \eqref{eq:define2}, which increases exponentially with the threshold $c$. One key step is to establish a  high-probability bound on the sum of two MMDs between  $\widetilde{\pi}_P'$ and $F^t_{\widetilde{X}'}$ ($F^t_{\widetilde{Y}'}$) (see Proposition \ref{prop:prob2} in Appendix \ref{proof:proposition2}). The main challenge is that the McDiarmid’s inequality \cite{havet2020quantitative} cannot be applied to HMMs. 
Let $h'=\sigma'-2a_P$. Due to the fact that  $a_P=\mathcal O\left(\frac{1}{\sqrt{m}}\right)$,  there always exists  a large   $m$ that $h'>0$. Define a function $\phi'$ as follows:
$\phi'(q')= \frac{\sqrt{\pi}}{\sqrt{\Gamma'}}q' \exp \Big({-q'h'+\frac{q'^2}{4\Gamma'}}\Big)$, where $\Gamma'>0$ is defined in Proposition \ref{prop:prob2}. We choose a constant $q'>0$ such that $\phi'(q') \le\frac{1}{2}$. Note that $\phi'(q')$ is a continuous function of $q'$, with $\phi'(0)=0$, $\phi'(q') \to \infty$ as $q' \to \infty$.
Therefore, there always exists such a value for $\phi'$.

\begin{theorem}\label{arl2}
The lower bound on the ARL in \eqref{eq:define2} increases exponentially with the threshold $c$:
$
    \textup{ARL}(T(c))\ge m\exp(q'c).
$
\end{theorem}
The proof is similar to the one for Theorem \ref{thm:arl} and is omitted.

\section{Numerical Results}
\begin{figure*}[!ht]
\begin{multicols}{3}
\includegraphics[width=0.85\linewidth]{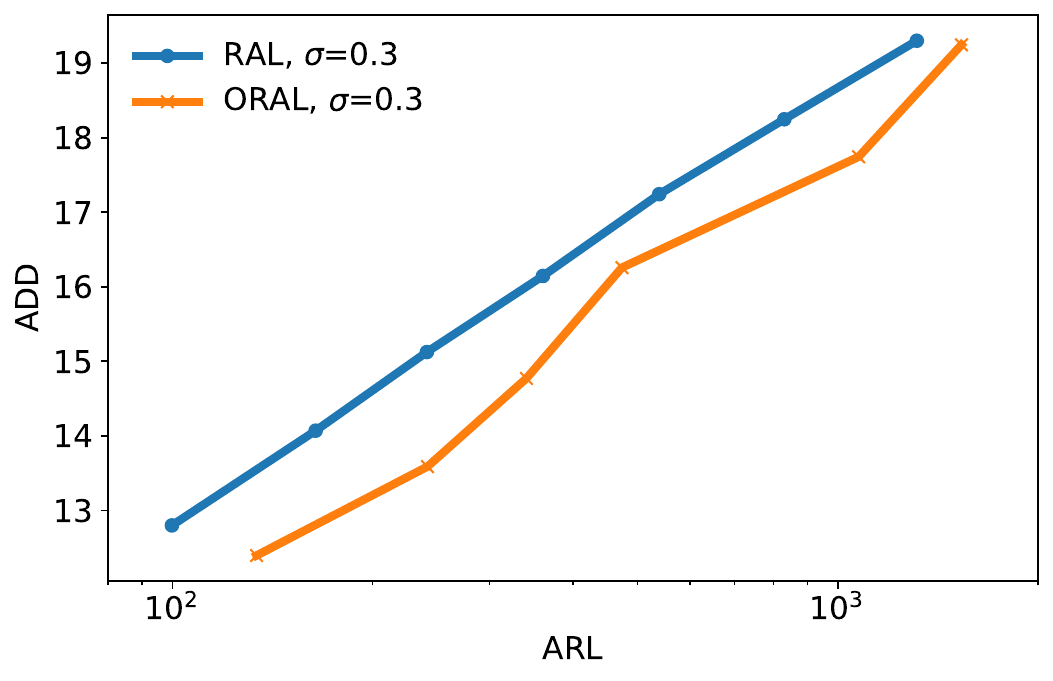}\par\caption{Comparison of the two algorithms for synthetic data: $\sigma=0.3$.}\label{fig:1}
\includegraphics[width=0.85\linewidth]{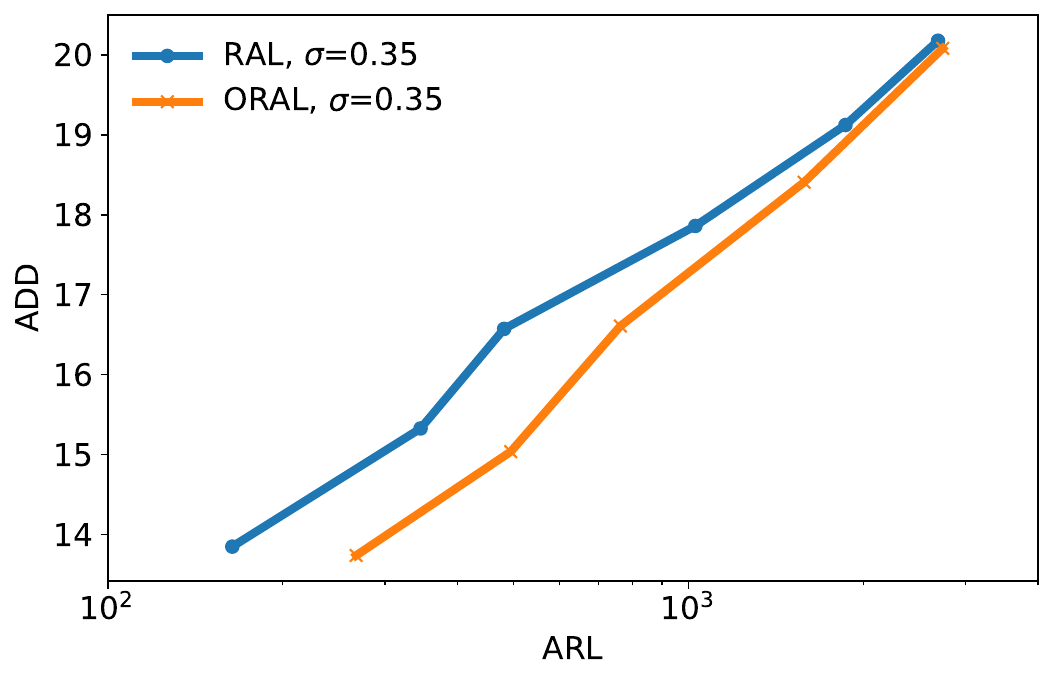}\par\caption{Comparison of the two algorithms for synthetic data: $\sigma=0.35$.}\label{fig:2}
\includegraphics[width=0.85\linewidth]{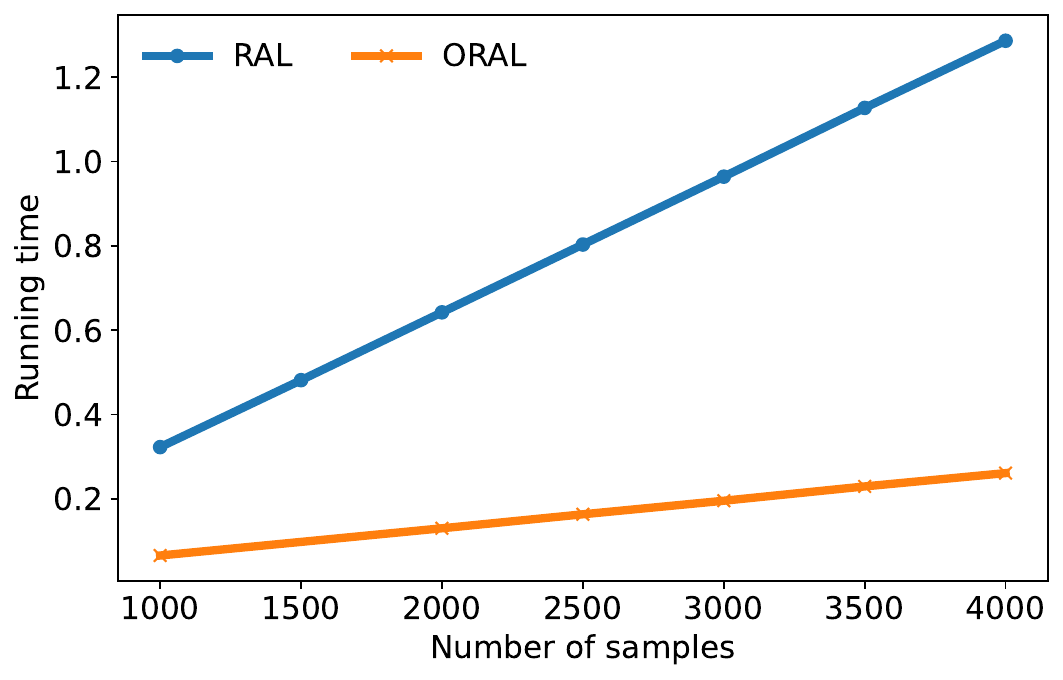}\par\caption{Number of samples v.s.  running time for synthetic data.}\label{fig:3}
\end{multicols}\vspace{-0.3cm}
\end{figure*}

\subsection{Synthetic Data}
We refer the algorithm introduced in \cite{chen2022change} as RAL (oveRlApping bLocks) and the proposed algorithm in this paper as ORAL (nOn-oveRlApping bLocks). We then compare their  performance.
We consider a case where the transition kernel of a Markov model changes from $ P=[0.2, 0.7,0.1; 0.9,0.0,0.1;0.2,0.8,0.0]^\top$
to $Q=[0.5, 0.5,0.0; 0.0,0.5,0.5;0.2,0.3,0.5]^\top$.
We utilize  the Gaussian kernel function $k(x,y)=\exp\big(-{\beta (x-y)^2}\big)$, with the bandwidth parameter $\beta$.  We choose $m=10$ and test  two different values of $\sigma$, $0.3$ and $0.35$ respectively. We select $\beta=\frac{1}{m-1}$ \cite{scikit-learn } which attains the best tradeoff between ADD and ARL. To compare the ADD and the ARL, in Figs \ref{fig:1} and \ref{fig:2}, we plot the ADD as a function of the logarithm of ARL by adjusting the threshold. Additionally, we provide a comparison of the computational complexity in Fig \ref{fig:3}, showing the running time in relation to the overall number of  samples (on Intel W-2295 CPU).

\begin{figure*}[!ht]
\centering 
\includegraphics[height=0.45\textwidth,width=0.9\textwidth]{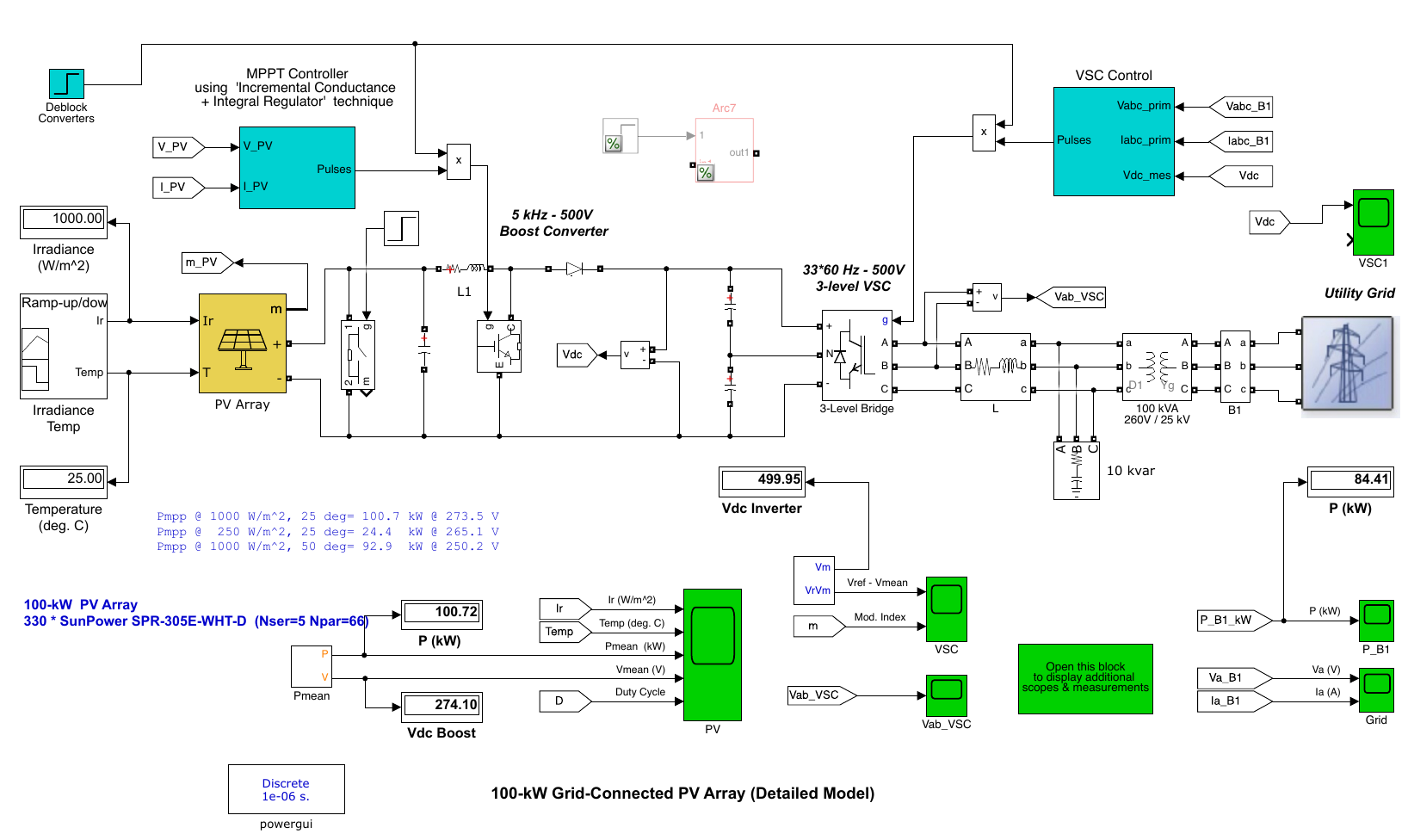}  
\caption{Simulation system for Line-Line fault in PV.} 
\label{fig:5}
\end{figure*}

Figs \ref{fig:1} and \ref{fig:2} clearly demonstrate that our approach outperforms the approach in \cite{chen2022change}, i.e., for a given ARL, our method requires fewer samples to detect the change. Furthermore, both methods exhibit a linear relationship between the ADD and the logarithm of ARL, which  aligns with our WADD upper bounds derived in this paper.
Fig \ref{fig:3} demonstrates the computationally efficiency of our algorithm. 

\subsection{Fault Detection in DC Microgrid}
\begin{figure}[!ht]
\centering 
\includegraphics[width=0.5\textwidth]{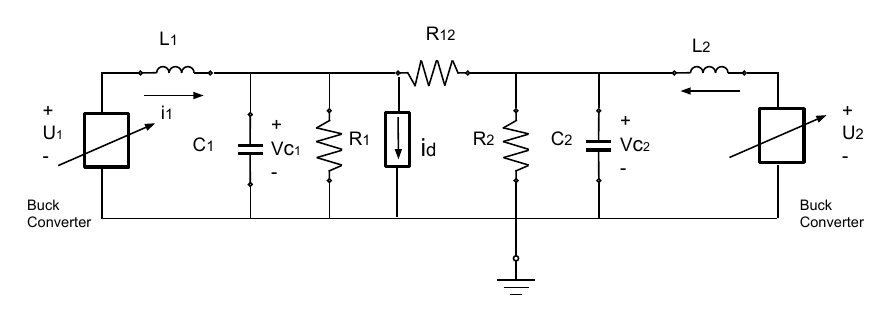} 
\caption{ Distribution changes in DC Microgrid.} 
\label{fig:4}
\end{figure}
We consider a discrete state space model of a DC microgrid. The circuit for this model is shown in Fig \ref{fig:4}. There are two buck converters and the reference voltage $V_{ref}=270$ V for both. For generator 1, set the inductance $L_1=1$ mH, the capacitance $C_1=2$ mF, the parallel resistor $R_1=1\;\Omega$ and the droop gain $rd_1$ of the buck converter is $0.15$. For generator 2, set the inductance $L_2=2$ mH, the capacitance $C_2=10$ mF, the parallel resistor $R_2=1000\;\Omega$, the droop gain $rd_2=0.1$.  

Due to a certain current source load or a line-to-ground fault, there is a current source $i_d$ in parallel with $R_1$ and we use $W$ to denote this disturbance. The change of this disturbance indicates the presence of a fault in either the load itself or a line to ground fault. According to the circuit, we derive the following discretized state space equations for this  system:
\begin{flalign}
      X_{t+1}&=A_dX_t+B_d V_{ref}+D_d W_t,\nonumber
\end{flalign}
where $X_t=[i_1^t,V_{C_1}^t,V_{ref}-V_{C_1}^t,i_2^t,V_{C_2}^t,V_{ref}-V_{C_2}^t]^\top$ is the system state and $W_t$ is the disturbance for the $t$-th time period and the time step is $1$ ms. Then we detect the different changes of the disturbance $W_t$.

Set $m=5, \sigma=0.45$ and $0.5, 0.55,$ respectively. We choose the Gaussian kernel function and  we pick $\beta=\frac{1}{4}$ that achieves the best ADD and ARL tradeoff.
Firstly, we only change the mean of the distribution of disturbance. For the pre-change, $W_k$ follows $\mathcal N(0,1)$. After the change point, the distribution of the process fault changes into $\mathcal N(5,1)$.  In Table \ref{tab:label1}, we provide the ADD and ARL with a group of thresholds.

We then change the variance of the distribution of disturbance. For the pre-change, $W_k$ follows $\mathcal N(0,1)$. After the change point, the distribution of the process fault changes into $\mathcal N(0,4)$. The results can be found in Table \ref{tab:label2}.
Finally, we change both the mean and variance.  For the pre-change, $W_k$ follows $\mathcal N(0,1)$. After the change point, the distribution of the process fault changes into $\mathcal N(5,4)$.   In Table \ref{tab:label3}, we provide the ADD and ARL with a group of thresholds.
From the above Tables,  it can be seen that for both algorithms, the ADD grows with the log of ARL linearly. Moreover, our method outperforms the method in \cite{chen2022change}, i.e., for the same level of ARL, our method needs less samples to raise an alarm.
\begin{table}[!ht]
    \centering
    \begin{tabular}{p{0.2\linewidth} | p{0.05\linewidth}| p{0.08\linewidth}| p{0.08\linewidth}| p{0.08\linewidth}| p{0.08\linewidth}}
     \hline
\multirow{2}{5em}{RAL,$\sigma$=0.45} & ADD &7.29&	7.44&	7.57&	7.73 \\ 
& ARL & 13465&	21489&	26889&	37432 \\ 
\hline
\multirow{2}{5em}{ORAL,$\sigma$=0.45 } & ADD &5.80&6.01&6.18& 6.62 \\ 
& ARL & 13839&19697&23361& 42792\\ 

\hline
\multirow{2}{5em}{RAL,$\sigma$=0.5 } & ADD & 7.30&	7.45&	7.58&	7.73\\ 
& ARL & 13414&	21668&	26865&	37737\\ 
\hline
\multirow{2}{5em}{ORAL,$\sigma$=0.5 } & ADD & 5.79&	6.01&	6.20&	6.64 \\ 
& ARL & 13881&	19614&	23255&	42684\\ 

\hline
\multirow{2}{5em}{RAL,$\sigma$=0.55 } & ADD & 7.44&	7.66&	7.97&	8.12 \\ 
& ARL & 57148&	79165&	138884&	185535\\ 
\hline
\multirow{2}{5em}{ORAL,$\sigma$=0.55 } & ADD & 6.20&	6.51&	6.76&	7.26\\ 
& ARL & 43347&	76092&	90716&	161087\\ 
\hline
    \end{tabular}
    \caption{Change in mean.}
    \label{tab:label1}
\end{table}

\begin{table}[!ht]
    \centering
    \begin{tabular}{p{0.2\linewidth} | p{0.05\linewidth}| p{0.08\linewidth}| p{0.08\linewidth}| p{0.08\linewidth}| p{0.08\linewidth}}
     \hline
\multirow{2}{5em}{RAL,$\sigma$=0.45} & ADD &49.83&	53.98&	57.63&	60.43 \\ 
& ARL & 4799&	6609&	8744&	10469 \\ 
\hline
\multirow{2}{5em}{ORAL,$\sigma$=0.45 } & ADD &50.45&53.11&56.8& 60.01 \\ 
& ARL &5404&7222&9216& 11923\\ 

\hline
\multirow{2}{5em}{RAL,$\sigma$=0.5 } & ADD & 64.54&	71.03&	76.04&	81.09\\ 
& ARL & 15479&	22094&	27274&	33262\\ 
\hline
\multirow{2}{5em}{ORAL,$\sigma$=0.5 } & ADD & 60.00&	65.89&	69.85&	75.34 \\ 
& ARL & 13309&	19227&	22570&	30724\\ 

\hline
\multirow{2}{5em}{RAL,$\sigma$=0.55 } & ADD & 87.38&	96.74&	112.46&	120.71 \\ 
& ARL & 50650&	82322&	177690&	245290\\ 
\hline
\multirow{2}{5em}{ORAL,$\sigma$=0.55 } & ADD & 80.08&	88.25&	96.98&	111.58\\ 
& ARL & 43726&	78065&	100750&	191223\\ 
\hline
    \end{tabular}
    \caption{Change in variance.}
    \label{tab:label2}
\end{table}

\begin{table}[!ht]
    \centering
    \begin{tabular}{p{0.2\linewidth} | p{0.05\linewidth}| p{0.08\linewidth}| p{0.08\linewidth}| p{0.08\linewidth}| p{0.08\linewidth}}
     \hline
\multirow{2}{5em}{RAL,$\sigma$=0.45} & ADD &7.05&	7.13&	7.22&	7.37 \\ 
& ARL & 4576&	7623&	8513&	9743 \\ 
\hline
\multirow{2}{5em}{ORAL,$\sigma$=0.45 } & ADD &5.10&5.16&5.25& 5.52 \\ 
& ARL & 4276&6298&8294& 11432\\ 

\hline
\multirow{2}{5em}{RAL,$\sigma$=0.5 } & ADD & 7.14&	7.31&	7.47&	7.67\\ 
& ARL & 16348&	22836&	29068&	35411\\ 
\hline
\multirow{2}{5em}{ORAL,$\sigma$=0.5 } & ADD & 5.25&	5.44&	5.62&	6.12 \\ 
& ARL & 13682&	18080&	24296&	39112\\ 

\hline
\multirow{2}{5em}{RAL,$\sigma$=0.55 } & ADD & 7.29&	7.54&	7.92&	8.03 \\ 
& ARL & 51673&	83208&	151070&	211081\\ 
\hline
\multirow{2}{5em}{ORAL,$\sigma$=0.55 } & ADD & 5.61&	5.97&	6.25&	6.99\\ 
& ARL & 45251&	65933&	98174&	157802\\ 
\hline
    \end{tabular}
    \caption{Change in both mean and variance.}
    \label{tab:label3}
\end{table}
\subsection{Fault Detection in Photovoltaic System}
In this section, we implement the proposed algorithm to a
simulated Photovoltaic (PV) system to detect the line-to-line fault. An auto-regressive process can be used to model faulty signal and noise in the PV system \cite{chen2016quickest}, which is an HMM. In Fig \ref{fig:5} we show our simulated PV system in Matlab by applying the toolbox SimPowerSystems. In our system, there is one 100-kW PV panel array with 5 × 66 PV
modules as the input, a boost converter, a three-level voltage source converter (VSC), a 25-kV grid
as the output and the simulation time step is $1\;\mu\text{s}$. We choose the maximum power point tracking (MPPT) with Incremental Conductance and Integral Regulator technique as our system controller.
We consider a line-line fault, which is simulated by a timed breaker.  During the change, the breaker is closed and the positive output of the PV array is connected to the negative output. We set the breaker resistance to $5\;\Omega$.
Without requiring the knowledge of current or power, we only need the output voltage of the PV system $V_{dc}$ to run our algorithm. Set $m=15, \sigma=0.04$. We choose the Gaussian kernel function and  we pick $\beta=\frac{1}{14}$.

We compare our approach with a model-based change detection algorithm \cite{gustafsson2000adaptive,liu2017sensor}, where adaptive filters are combined with the CuSum algorithm to detect the change. We design a Kalman filter for our PV system. Then the output residual signals should follow Gaussian distributions with zero mean. The change in the system will lead to a change in the mean and variance of the Gaussian distributions. The construction of the Kalman filters needs the entire knowledge of the PV circuits, which is hard to obtain in the data-driven setting. Hence, we measure the mean and variance of the residual signals with the unmatched Kalman filter, where only one parameter, the input capacitance $C_i$, is different from the ideal model. Then a CuSum algorithm is applied.
\begin{table}[!ht]
    \centering
    \begin{tabular}{p{0.2\linewidth} | p{0.05\linewidth}| p{0.08\linewidth}| p{0.08\linewidth}| p{0.08\linewidth}| p{0.08\linewidth}| p{0.08\linewidth}}
     \hline
\multirow{2}{5em}{Model Based,$C_i$=1} & ADD & 20.03&	23.74&	24.53&	26.37&	30.51 \\ 
& ARL & 224.59&	257.9&	265.04&	277.36&	319.36 \\ 
\hline
\multirow{2}{5em}{Model Based,$C_i$=0.5 } & ADD & 19.91&	23.47&	25.51&	28.27&	29.54 \\ 
& ARL & 76.05&	130.49&	145.29&	231.98&	285.81 \\ 
\hline
\multirow{2}{5em}{ORAL } & ADD & 20.55&	22.36&	23.55&	27.45&	31.05 \\ 
& ARL & 4485.85&	5014.53&	5228.55&	5589.60&	5840.61\\ 
\hline
    \end{tabular}
    \caption{Model-based algorithm v.s. our algorithm.}
    \label{tab:my_label}
\end{table}

In Table \ref{tab:my_label}, we provide the ADD and ARL with different thresholds. It shows that for the same level of ADD, our method achieves a larger ARL, which demonstrates that our method outperforms the model-based method with unmatched system information.

\section{Conclusion}

In this paper, we investigated the data-driven quickest change detection problem, focusing on both Markov models and HMMs. We developed kernel-based detection algorithms and analyzed the bias in MMD estimates for both Markov models and HMMs. Moreover, we established theoretical lower bound on ARL and upper bound on WADD. We demonstrated that the WADD is at most in the logarithm of the ARL.  
Combined with the universal lower bound on the WADD in \cite{lai1998information}, our approach is optimal at order-level. Furthermore, in the case of Markov models, compared with existing state-of-the-art study in the same setting, our theoretical bounds are more stringent and our algorithms are computationally efficient. Our algorithm is the first data-driven one for QCD in
HMMs with theoretical performance guarantees. Simulation results demonstrate the performance of our proposed algorithms.


\appendix

\subsection{Proposition \ref{proposition:mean} and Proof}\label{proof:proposition 1}
Recall that $ \widetilde{\pi}_P$ ( $ \widetilde{\pi}_Q$) and $ F^t_{\tilde{X}}$ are the invariant measure and empirical measure of $\{\widetilde{X}_t\}_{t=1}^\infty$ under $\mathbb P_{\infty}$ ($\mathbb P_1$), respectively. The proof of Theorem \ref{thm:add} uses the following upper bounds of the expectation of their MMD.
\begin{proposition} \label{proposition:mean}
$\forall x_{} \in  \mathcal X,$  
    $\mathbb E_{\infty}\Big[D\Big( F^t_{\widetilde{X}}, {\widetilde{\pi}_P}\Big)|X_{mt}=x_{}\Big]\le a_P$ and $\mathbb E_{1}\Big[D\Big( F^t_{\widetilde{X}}, {\widetilde{\pi}_Q}\Big)|X_{mt}=x_{}\Big]\le a_Q$.
\end{proposition}

\begin{proof}
 Recall that the samples in this paper are non-i.i.d., we define the coefficient  
$\rho_{\infty}(x_0,i,j)=\Big |\mathbb E_{\infty}\Big[\langle k\Big(\widetilde{X}_i,\cdot\Big)-\mu_{\widetilde{\pi}_P},k\Big(\widetilde{X}_j,\cdot\Big)-\mu_{\widetilde{\pi}_P}\rangle_{\mathcal{H}_k}|{X}_0={x}_0\Big] \Big|
$
under $\mathbb P_{\infty}$  and 
$\rho_{1}(x_0,i,j)=\Big |\mathbb E_{1}\Big[\langle k\Big(\widetilde{X}_i,\cdot\Big)-\mu_{\widetilde{\pi}_Q},k\Big(\widetilde{X}_j,\cdot\Big)-\mu_{\widetilde{\pi}_Q}\rangle_{\mathcal{H}_k}|{X}_0={x}_0\Big] \Big|
$ 
 under $\mathbb P_1$. 
\begin{lemma}\label{lemma:rho}
 Under $\mathbb P_{\infty}$ and $ \mathbb P_1$ respectively, for $\forall{x}_0 \in \mathcal X, j>i>0$, we have 
$
 \rho_{\infty}({x}_0,i,j) \le 2R_P\lambda_P^{j-i-1},
\rho_{1}({x}_0,i,j)\le 2R_Q\lambda_Q^{j-i-1}.
$
\end{lemma}
The proof of Lemma \ref{lemma:rho} can be found in Appendix \ref{proof:lemma2}. 

Note the kernel $k$ is bounded. Then, 
$
    \mathbb E_{\infty}\Big[\|k(\widetilde{X}_i,\cdot)-\mu_{ \widetilde{\pi}_P}\|_{\mathcal{H}_k}^2\Big|X_0=x_0\Big]\le 2.
$
Therefore, we have that
\begin{flalign}
    \mathbb E&_{\infty}\Big[D( F^t_{\widetilde X}, {\widetilde{\pi}_P})^2|X_{mt}=x_{mt}\Big] \nonumber\\
    =& \frac{1}{(m-1)^2}\mathbb E_{\infty}\Bigg[\sum_{i=1}^{m-1}\Big\|k(\widetilde{X}_{mt+i},\cdot)-\mu_{ \widetilde{\pi}_P}\Big\|_{\mathcal{H}_k}^2 |X_{mt}=x_{mt}\Bigg]\nonumber\\
    &+2\sum_{mt<i<j< mt+m}\mathbb E_{\infty}\Big[\Big\langle k(\widetilde{X}_i,\cdot)-\mu_{ \widetilde{\pi}_P},\nonumber\\
    &k(\widetilde{X}_j,\cdot)-\mu_{ \widetilde{\pi}_P}\Big\rangle_{\mathcal{H}_k} |X_{mt}=x_{mt}\Big]\nonumber\\
    \le& \frac{1}{(m-1)^2}\Bigg(2(m-1)+2\sum_{mt<i<j< mt+m}\rho_{\infty}(x_{mt},i,j)\Bigg)\nonumber\\
    \le& \frac{1}{m-1}\Bigg(2+4\frac{R_P}{1-\lambda_P}\Bigg).
\end{flalign}
Moreover, $\mathbb E_{\infty}\left[D( F^t_{\widetilde X},  \widetilde{\pi}_P)|X_{mt}=x_{mt}\right]\le  \sqrt{\mathbb E_{\infty}[D( F^t_{\widetilde X},  \widetilde{\pi}_P)^2|X_{mt}=x_{mt}]}\le a_P$. Since $a_P$ is a constant, there exists an upper bound on the expectation of MMD between the empirical measure and invariant measure for each block. The proof under $\mathbb P_1$ follows similarly.
\end{proof}

\subsection{Proof of Lemma \ref{lemma:rho}}\label{proof:lemma2}

We provide the proof under $\mathbb P_{\infty}$. The proof under $\mathbb P_1$ can be derived in the same way. It can be shown that
\begin{flalign}
    \rho&_{\infty}(x_0,i,j)
    =\Big|\int\int \mathbb P_{\infty}(d\widetilde{x}_j|\widetilde{X}_i=\widetilde{x}_i,X_0=x_0)\nonumber\\
    &\times \mathbb P_{\infty}(d\widetilde{x}_i|X_0=x_0)k(\widetilde{x}_i,\widetilde{x}_j) \nonumber\\
    &-\int\int \mathbb P_{\infty}(d\widetilde{x}_i|X_0=x_0) \widetilde{\pi}_P(d\widetilde{x})k(\widetilde{x}_i,\widetilde{x}) \nonumber\\
    &-\int\int \mathbb P_{\infty}(d\widetilde{x}_j|X_0=x_0) \widetilde{\pi}_P(d\widetilde x)k(\widetilde{x}_j,\widetilde{x}) \nonumber\\
    &+\int\int  \widetilde{\pi}_P(d\widetilde{x}) \widetilde{\pi}_P(d\widehat{x})k(\widetilde{x},\widehat{x}) \Big|\nonumber\\
    \overset{(a)}{\le}&\Big|\int\int \mathbb P_{\infty}(d\widetilde{x}_i|X_0=x_0)k(\widetilde{x}_i,\widetilde{x}_j)\nonumber \\
    &\times \big(\mathbb P_{\infty}(d\widetilde{x}_j|\widetilde{X}_i=\widetilde{x}_i,X_0=x_0)- \widetilde{\pi}_P(d\widetilde{x}_j)\big)  \Big|\nonumber\\
    &+\Big|\int\int  \widetilde{\pi}_P(d\widetilde{x}) k(\widetilde{x},\widetilde{x}_j)\nonumber\\
    & \times \big(\mathbb P_{\infty}(d\widetilde{x}_j|X_0=x_0)- \widetilde{\pi}_P(d\widetilde{x}_j)\big)\Big|\nonumber\\
    \overset{(b)}{\le}& 2R_P\lambda_P^{j-i-1},
\end{flalign}
where (a) is due to the absolute value inequality and (b) is due to the uniformly ergodicity.

\subsection{Proposition \ref{prop:prob} and Proof}\label{proof:prob}
Define  
$
     \zeta=\inf\{i\ge 1:2R_P\lambda_P^i<1\},$ $
     \eta=2R_P\lambda_P^\zeta, $ and $
     \chi=\eta^{-(\zeta-1)/\zeta}\Big\{1-\eta^{1/\zeta}/2\Big\}^{-1}. 
$
 Let 
$
     u=\Big\{1+\eta^{1/\zeta}/2\Big\}^{-1},$ $
     M=\frac{2R_P-1}{1-(2R_P)^{-1/\zeta}}\chi+\frac{1-\eta^{-1}}{1-\eta^{-1/\zeta}}+2\Big(1+\eta^{1/\zeta}\Big)^{-1}\eta^{-(\zeta-a)/\zeta}.
$
Then $\Gamma$ is defined as  
\begin{align*}
    \Gamma= \frac{(1-\max(\lambda_P,u^{-1/4}))^2}{128mR_P\chi}\Big(\frac{5}{\log(u)}+8MR_P\chi\Big)^{-1}.
\end{align*}

We derive a  proposition that provides a high-probability bound on the sum of two MMDs between the invariant measure and the empirical measure. 
\begin{proposition} \label{prop:prob}
For any $x_{mt}$ ,$y_{mt} \in \mathcal X$ and $\delta>0$, 
$
    \mathbb P_{\infty}\Big(D( F^t_{\widetilde X}, \widetilde{\pi}_P)+D( F^t_{\widetilde Y}, \widetilde{\pi}_P)\le \sqrt{\frac{\log(\frac{1}{\delta})}{\Gamma}} 
    +2a_P\big|X_{mt}=x_{mt},Y_{mt}=y_{mt}\Big) \ge 1-\delta. 
$
\end{proposition}

\begin{proof}
For any $t\ge0$ and $m>1$, we define a function
\begin{flalign}
    g&\Big((\widetilde{X}_{mt+1},\widetilde{Y}_{mt+1}),\ldots, (\widetilde{X}_{mt+m-1},\widetilde{Y}_{mt+m-1})\Big)\nonumber\\
    =&\frac{1}{m-1}\Bigg\|\sum_{i=1}^{m-1}(\mu_{\delta_{{\widetilde{X}}_{mt+i}}}-\mu_{ \widetilde{\pi}_P})\Bigg\|_{\mathcal{H}_k} \nonumber\\
    &+\frac{1}{m-1}\Bigg\|\sum_{i=1}^{m-1}(\mu_{\delta_{\widetilde{Y}_{mt+i}}}-\mu_{ \widetilde{\pi}_P})\Bigg\|_{\mathcal{H}_k}.
\end{flalign}
When there is no change, the Markov chains $\{\widetilde{X}_n\}_{n=1}^\infty$ and $\{\widetilde{Y}_n\}_{n=1}^\infty$ are uniformly ergodic. Then it can be shown that the Markov chain $\{(\widetilde{X}_n,\widetilde{Y}_n\}_{n=1}^\infty$ is uniformly ergodic.  For any $\widetilde{x}_i, \widetilde{y}_i$,  $A,B \subseteq  \widetilde{\mathcal X}$ and $i<j$, it can be easily shown that 
$
\Big|\mathbb P_{\infty}\big(\widetilde{X}_j\in A,\widetilde{Y}_j \in B\big|\widetilde{X}_i=\widetilde{x}_i,\widetilde{Y}_i=\widetilde{y}_i\big) 
-\mathbb P_{\infty}\big(\widetilde{X}_j\in A,\widetilde{Y}_j\in B\big) \Big|
\le  2R_P\lambda_P^{j-i}. 
$

By \cite[Theorem 3.1]{havet2020quantitative}, which is a generalization of the McDiarmid’s inequality to uniformly ergodic Markov chains, it can be shown that for any $\omega>0$,
\begin{flalign}
    \mathbb P&_{\infty}\Big(D( F^t_{\widetilde X}, \widetilde{\pi}_P)+D(F^t_{\widetilde Y}, \widetilde{\pi}_P)-\mathbb E_{\infty}[D(F^t_{\widetilde X}, \widetilde{\pi}_P) \nonumber\\
    &+D(F^t_{\widetilde Y}, \widetilde{\pi}_P)] \ge \omega
    \big|X_{mt}=x_{mt},Y_{mt}=y_{mt}\Big) \nonumber\\
    =&\mathbb P_{\infty}\Big(g\big((\widetilde{X}_{mt+1},\widetilde{Y}_{mt+1}),\ldots, (\widetilde{X}_{mt+m-1},\widetilde{Y}_{mt+m-1})\big)\nonumber\\
    &-\mathbb E_{\infty}\Big[g\big((\widetilde{X}_{mt+1},\widetilde{Y}_{mt+1}),\ldots, (\widetilde{X}_{mt+m-1},\widetilde{Y}_{mt+m-1})\big)\Big]\nonumber\\
    &\ge \omega\big|X_{mt}=x_{mt},Y_{mt}=y_{mt}\Big)\nonumber\\
    \le& \exp\Big(-\Gamma \omega^2\Big).
\end{flalign}
Let $\omega=\sqrt{\frac{\log(\frac{1}{\delta})}{\Gamma}}$. The result  then follows from 
Proposition \ref{proposition:mean}.
\end{proof}
\subsection{Sufficient Condition to Guarantee Detectability} \label{proof:l1}
\begin{lemma}\label{assume}
For any $A,B_1,\ldots,B_{i-1}\subseteq \mathcal X$, if the conditional distributions $\pi_P'(X_{i}'\in A|X_1'\in B_1,\ldots,X_{i-1}'\in B_{i-1})\neq  \pi_Q'(X_{i}'\in A|X_1'\in B_1,\ldots,X_{i-1}'\in B_{i-1})$, then we have 
\begin{flalign}
&\pi_P'(X_1'\in B_1,\ldots,X_{i-1}'\in B_{i-1},X_{i}'\in A)\nonumber\\
&\neq  \pi_Q'(X_1'\in B_1,\ldots,X_{i-1}'\in B_{i-1},X_{i}'\in A).
\end{flalign}
\end{lemma}
\begin{proof}[Proof]
Firstly, we have that for any $A,B_1,\ldots,B_{i-1}\subseteq \mathcal X$, 
\begin{flalign}
    &\pi_P'(X_{t+i-1}'\in A|X_t'\in B_1,\ldots,X_{t+i-2}'\in B_{i-1})\nonumber\\
    &=\frac{\pi_P'(X_t'\in B_1,\ldots,X_{t+i-2}'\in B_{i-1},X_{t+i-1}'\in A)}{\pi_P'(X_t'\in B_1,\ldots,X_{t+i-2}'\in B_{i-1})}.\label{eq:pfl3}
\end{flalign}
If $
    \pi_P'(X_t'\in B_1,\ldots,X_{t+i-2}'\in B_{i-1})
    \neq \pi_Q'(X_t'\in B_1,\ldots,X_{t+i-2}'\in B_{i-1})
$
and $A=\mathcal X$, it follows that
\begin{flalign}\label{eq:25}
    &\pi_P'(X_t'\in B_1,\ldots,X_{t+i-2}'\in B_{i-1},X_{t+i-1}'\in A)\nonumber\\
    &\neq \pi_Q'(X_t'\in B_1,\ldots,X_{t+i-2}'\in B_{i-1},X_{t+i-1}'\in A).
\end{flalign}
If $\pi_P'(X_t'\in B_1,\ldots,X_{t+i-2}'\in B_{i-1})= \pi_Q'(X_t'\in B_1,\ldots,X_{t+i-2}'\in B_{i-1}),$
 we can still show that \eqref{eq:25} holds.
This is because  from \eqref{eq:pfl3}
$
      \pi_P'(X_{t+i-1}'\in A|X_t'\in B_1,\ldots,X_{t+i-2}'\in B_{i-1}) 
      \neq  \pi_Q'(X_{t+i-1}'\in A|X_t'\in B_1,\ldots,X_{t+i-2}'\in B_{i-1}).$
\end{proof}

\subsection{Proposition \ref{proposition:mean2} and Proof}
\label{proof:lemma3}
\begin{proposition} \label{proposition:mean2}
$\forall x_{} \in  \mathcal X,$  
    $\mathbb E_{\infty}\Big[D\Big( F^t_{\widetilde{X}'}, {\widetilde{\pi}_P'}\Big)|X_{mt}=x_{}\Big]\le a_P$ and $\mathbb E_{1}\Big[D\Big( F^t_{\widetilde{X}'}, {\widetilde{\pi}_Q'}\Big)|X_{mt}=x_{}\Big]\le a_Q$.
\end{proposition}

For HMMs, we define the coefficients
$
\rho_{\infty}'(x_0,i,j)
=\Big |\mathbb E_{\infty}\Big[\langle k\Big(\widetilde{X}_i',\cdot\Big)-\mu_{\pi_P'},k\Big(\widetilde{X}'_j,\cdot\Big)-\mu_{\pi_P'}\rangle_{\mathcal{H}_k}|{X}_0={x}_0\Big] \Big|
$
under $\mathbb P_{\infty}$ and $
\rho_{1}'(x_0,i,j)=\Big |\mathbb E_1\Big[\langle k\Big(\widetilde{X}_i',\cdot\Big)-\mu_{\pi_Q'},k\Big(\widetilde{X}'_j,\cdot\Big)-\mu_{\pi_Q'}\rangle_{\mathcal{H}_k}|{X}_0={x}_0\Big] \Big|
$
under $\mathbb P_1$. 
\begin{lemma}\label{lemma:rho2}
 Under $\mathbb P_{\infty}$ and $ \mathbb P_1$, respectively, for $\forall{x}_0 \in \mathcal X, j>i>0$, we have 
$
 \rho_{\infty}'({x}_0,i,j) \le 2R_P\lambda_P^{j-i-1},$ and $
\rho_{1}'({x}_0,i,j)\le 2R_Q\lambda_Q^{j-i-1}.
$
\end{lemma}
Based on Lemma \ref{lemma:rho2}, the proof of Proposition \ref{proposition:mean2} follows the proof of Proposition \ref{proposition:mean} in Appendix \ref{proof:proposition 1} thus is omitted here.
We then provide the proof of the first inequality in Lemma \ref{lemma:rho2}. The proof of the second inequality can be derived via the same process.
For $i<j$ and $\widetilde A\subseteq \mathcal X \times \mathcal X$, we have that
\begin{flalign}
    \mathbb P_{\infty}&\left(\widetilde{x}_j'\in \widetilde A |\widetilde{X}_i'=\widetilde{x}_i',X_0=x_0\right)- \widetilde{\pi}_P'\Big(\widetilde{x}_j'\in \widetilde A\Big)\nonumber\\
    =& \int_{x_j}\int_{x_{i+1}} \mathbb P_{\infty}\Big(\widetilde{x}_j'\in \widetilde A|X_j=x_j\Big)\nonumber\\
   & \times \mathbb P_{\infty}(dx_{i+1}|\widetilde{X}_i'=\widetilde{x}_i', X_0=x_0)\nonumber\\
    & \times \Big(\mathbb P_{\infty}(dx_j|X_{i+1}=x_{i+1})-\pi_P(dx_j)\Big).
\end{flalign}
Due to the uniform ergodicity of $\{X_t\}_{t=0}^\infty$, it follows that
$
\Big|\mathbb P_{\infty}\left(\widetilde{x}_j'\in \widetilde A|\widetilde{X}_i'=\widetilde{x}_i',X_0=x_0\right)- \widetilde{\pi}_P'\Big(\widetilde{x}_j'\in \widetilde A\Big)\Big|\le R_P\lambda_P^{j-i-1}. 
$
Thus, it follows that 
\begin{flalign}
    \rho&_{\infty}'(x_0,i,j)
    =\Big|\mathbb E_{\infty}\Big[k\Big(\widetilde{X}_i',\widetilde{X}_j'\Big)|X_0=x_0\Big]\nonumber\\
    &-\mathbb E_{\widetilde{X}'\sim  \widetilde{\pi}_P'}\Big[k\Big(\widetilde{X}_i',\widetilde{X}'\Big)|X_0=x_0\Big] \nonumber \\
    &-\mathbb E_{\widetilde{X}'\sim \widetilde{\pi}_P'}\Big[k\Big(\widetilde{X}_j',\widetilde{X}'\Big)|X_0=x_0\Big]\nonumber\\
    &+\mathbb E_{\widetilde{X}',\widehat{X}'\sim  \widetilde{\pi}_P'}\Big[k\left(\widetilde{X}',\widehat{X}'\right)\Big]\Big|\nonumber\\
    \le&\Big|\int\int \mathbb P_{\infty}\Big(
    d\widetilde{x}_i'|X_0=x_0\Big)k\big(\widetilde{x}_i',\widetilde{x}_j'\big)\nonumber\\
    &\times \Big(\mathbb P_{\infty}\Big(d\widetilde{x}_j'|\widetilde{X}_i'=\widetilde{x}_i',X_0=x_0\Big)- \widetilde{\pi}_P'(d\widetilde{x}_j')\Big)\Big|\nonumber\\
    &+\Big|\int\int  \widetilde{\pi}_P'(d\widetilde{x}')k\big(\widetilde{x}',\widetilde{x}_j'\big)\nonumber\\
    &\times \Big(\mathbb P_{\infty}\Big(d\widetilde{x}_j'|X_0=x_0\Big)- \pi_P'(d\widetilde{x}_j')\Big)
    \Big|\nonumber\\
    \le& 2R_P\lambda_P^{j-i-1}.
\end{flalign}

\subsection{Proposition \ref{prop:prob2} and Proof}
\label{proof:proposition2}

\begin{proposition} \label{prop:prob2}
For any $x_{}$, $y_{} \in \mathcal X$ and $\delta>0$, 
\begin{flalign}
    &\mathbb P_{\infty}\Big(D\big( F^t_{\widetilde X'}, \widetilde{\pi}_P'\big)+D\big( F^t_{\widetilde Y'}, \widetilde{\pi}_P'\big)\le \sqrt{\frac{\log(\frac{1}{\delta})}{\Gamma'}} \nonumber\\
    &+2a_P\Big|X_{mt}=x_{},Y_{mt}=y_{}\Big) \ge 1-2\delta, \label{eq:p4}
\end{flalign}
where $\Gamma'=\frac{\Gamma}{(1+\sqrt{2m\Gamma})^2}$ and $\Gamma>0$ is defined in Appendix \ref{proof:prob}.
\end{proposition}
\begin{proof}
For any $t\ge0$ and $m>1$, we define two functions
\begin{flalign}
    g&\Big(\left(\widetilde{X}_{mt+1}',\widetilde{Y}_{mt+1}'\right),\ldots, \Big(\widetilde{X}_{mt+m-1}',\widetilde{Y}_{mt+m-1}'\Big)\Big)\nonumber\\
    =&\frac{1}{m-1}\Bigg\|\sum_{i=1}^{m-1}(\mu_{\delta_{\tilde{X}'_{mt+i}}}-\mu_{ \widetilde{\pi}_P'})\Bigg\|_{\mathcal{H}_k} \nonumber\\
    &+\frac{1}{m-1}\Bigg\|\sum_{i=1}^{m-1}(\mu_{\delta_{\widetilde{Y}'_{mt+i}}}-\mu_{ \widetilde{\pi}_P'})\Bigg\|_{\mathcal{H}_k},\\
    f&\Big(\Big(\widetilde{X}_{mt+1},\widetilde{Y}_{mt+1}\Big),\ldots, \Big(\widetilde{X}_{mt+m-1},\widetilde{Y}_{mt+m-1}\Big)\Big)\nonumber\\
    =&\mathbb E_{\infty}\Big[g\Big(\Big(\widetilde{X}_{mt+1}',\widetilde{Y}_{mt+1}'\Big),\ldots, \Big(\widetilde{X}_{mt+m-1}',\widetilde{Y}_{mt+m-1}'\Big)\Big)\nonumber\\
    &\Big|\Big(\widetilde{X}_{mt+1},\widetilde{Y}_{mt+1}\Big),\ldots, \Big(\widetilde{X}_{mt+m-1},\widetilde{Y}_{mt+m-1}\Big)\Big].
\end{flalign}
We then have that 
\begin{flalign}
     &\mathbb E_{\infty}\Big[f\Big(\Big(\widetilde{X}_{mt+1},\widetilde{Y}_{mt+1}\Big),\ldots, \Big(\widetilde{X}_{mt+m-1},\widetilde{Y}_{mt+m-1}\Big)\Big)\Big]\nonumber\\
    =&\mathbb E_{\infty}\Big[g\Big(\Big(\widetilde{X}_{mt+1}',\widetilde{Y}_{mt+1}'\Big),\ldots, \Big(\widetilde{X}_{mt+m-1}',\widetilde{Y}_{mt+m-1}'\Big)\Big)\Big].\nonumber
\end{flalign}
For any $\omega>0$,
\begin{flalign}
    \mathbb P&_{\infty}\Big(D\Big( F^t_{\widetilde{X}'}, \widetilde{\pi}_P'\Big)+D\Big(F^t_{\widetilde Y'}, \widetilde{\pi}_P'\Big)-\mathbb E_{\infty}\Big[D\Big(F^t_{\widetilde X'}, \widetilde{\pi}_P'\Big) \nonumber\\
    &+D\Big(F^t_{\widetilde Y'}, \widetilde{\pi}_P'\Big)\Big] \ge \omega
    \big|X_{mt}=x_{mt},Y_{mt}=y_{mt}\Big) \nonumber\\
    =&\int\ldots\int \mathbb P_{\infty}\Big(d{x}_{mt+1},d{y}_{mt+1},\ldots,d{x}_{mt+m},d{y}_{mt+m}\nonumber\\
    &\big|X_{mt}=x_{mt},Y_{mt}=y_{mt}\Big)\times \Delta,\label{eq:pro4}
\end{flalign}
where 
\begin{flalign}
&\Delta=\mathbb P_{\infty}\Big(g\Big(\Big(\widetilde{X}'_{mt+1},\widetilde{Y}'_{mt+1}\Big),\ldots, \Big(\widetilde{X}'_{mt+m-1},\widetilde{Y}'_{mt+m-1}\Big)\Big)\nonumber\\
    &-f\big((\tilde{x}_{mt+1},\tilde{y}_{mt+1}),\ldots,(\tilde{x}_{mt+m-1},\tilde{y}_{mt+m-1})\big)\nonumber\\
    &+f\big((\tilde{x}_{mt+1},\tilde{y}_{mt+1}),\ldots,(\tilde{x}_{mt+m-1},\tilde{y}_{mt+m-1})\big)\nonumber\\
    -&\mathbb E_{\infty}\Big[g\Big(\Big(\widetilde{X}'_{mt+1},\widetilde{Y}'_{mt+1}\Big),\ldots, \Big(\widetilde{X}'_{mt+m-1},\widetilde{Y}'_{mt+m-1}\Big)\Big)\Big]\nonumber\\
    &\ge \omega\big|X_{mt}=x_{mt},Y_{mt}=y_{mt},\ldots,\nonumber\\
&X_{mt+m}=x_{mt+m},Y_{mt+m}=y_{mt+m}\Big).
\end{flalign}
For any $0\le l'\le\omega$, we have that 
$
    \Delta
    \le \Delta_1+\Delta_2,
$
    where 
    \begin{flalign}
&\Delta_1=
    \mathbb P_{\infty}\Big(g\Big(\Big(\widetilde{X}'_{mt+1},\widetilde{Y}'_{mt+1}\Big),\ldots, \Big(\widetilde{X}'_{mt+m-1},\widetilde{Y}'_{mt+m-1}\Big)\Big)\nonumber\\
    &-f\big((\tilde{x}_{mt+1},\tilde{y}_{mt+1}),\ldots,(\tilde{x}_{mt+m-1},\tilde{y}_{mt+m-1})\big)\nonumber\\
        &\ge l' \big|X_{mt}=x_{mt},Y_{mt}=y_{mt},\ldots,\nonumber\\
&X_{mt+m}=x_{mt+m},Y_{mt+m}=y_{mt+m}\Big)
\\
    &\Delta_2=\mathbb P_{\infty}\Big(
    f\big((\tilde{x}_{mt+1},\tilde{y}_{mt+1}),\ldots,(\tilde{x}_{mt+m-1},\tilde{y}_{mt+m-1})\big)\nonumber\\
    -&\mathbb E_{\infty}\Big[g\Big(\Big(\widetilde{X}'_{mt+1},\widetilde{Y}'_{mt+1}\Big),\ldots, \Big(\widetilde{X}'_{mt+m-1},\widetilde{Y}'_{mt+m-1}\Big)\Big)\Big]\nonumber\\
        &\ge \omega-l'\big|X_{mt}=x_{mt},Y_{mt}=y_{mt},\ldots,\nonumber\\
&X_{mt+m}=x_{mt+m},Y_{mt+m}=y_{mt+m}\Big)\label{ineq:devide}.
\end{flalign}
By \cite[Theorem 3.1]{havet2020quantitative}, which is a generalization of the McDiarmid’s inequality to uniformly ergodic Markov chains, we have that for any $w-l'\ge 0$,
$
    \Delta_2\le \exp(-\Gamma(\omega-l')^2).
$
Given any $X_{mt}=x_{mt},Y_{mt}=y_{mt},\ldots,X_{mt+m}=x_{mt+m},Y_{mt+m}=y_{mt+m}$,  $X_{mt+1}',Y_{mt+1}',\ldots, X_{mt+m}',Y_{mt+m}'$ are independent. Thus by McDiarmid’s inequality, we have that for any $l'\ge 0$ 
$
\Delta_1\le \exp\left(-\frac{l'^2}{2m}\right).
$
Let $l'=\frac{\sqrt{2m\Gamma}}{1+{\sqrt{2m\Gamma}}}\omega$, $\Gamma'=\frac{\Gamma}{(1+\sqrt{2m\Gamma})^2}$, we have that $\frac{l'^2}{2m}=\Gamma(\omega-l')^2$.
Then it follows that
\begin{flalign}
\mathbb P&_{\infty}\Big(D\Big( F^t_{\widetilde{X}'}, \widetilde{\pi}_P'\Big)+D\Big(F^t_{\widetilde Y'}, \widetilde{\pi}_P'\Big)-\mathbb E_{\infty}\Big[D\Big(F^t_{\widetilde X'}, \widetilde{\pi}_P'\Big) \nonumber\\
    &+D\Big(F^t_{\widetilde Y'}, \widetilde{\pi}_P'\Big)\Big] \ge \omega
\big|X_{mt}=x_{mt},Y_{mt}=y_{mt}\Big)\nonumber\\
\le & \exp\left(-\frac{l'^2}{2m}\right)+\exp(-\Gamma(\omega-l')^2)\nonumber\\
=&2\exp(-\Gamma'\omega^2).
\end{flalign}
Further applying Proposition \ref{proposition:mean2}
concludes the proof.
\end{proof}

\normalem
\bibliographystyle{ieeetr}
\bibliography{QCD,QCD_zou}

\end{document}